\documentclass{article}

\usepackage{PRIMEarxiv}

\usepackage[utf8]{inputenc} 
\usepackage[T1]{fontenc}    
\usepackage{hyperref}       
\usepackage{url}            
\usepackage{booktabs}       
\usepackage{amsfonts}       
\usepackage{nicefrac}       
\usepackage{microtype}      
\usepackage{lipsum}
\usepackage{fancyhdr}       
\usepackage{graphicx}       
\graphicspath{{media/}}     

\usepackage{epsfig}
\usepackage{amsmath}
\usepackage{amssymb}
\usepackage{color}
\usepackage{url}
\usepackage{stmaryrd}
\usepackage{algorithm}
\usepackage{algpseudocode}

\newcommand{\veryshortarrow}[1][3pt]{\mathrel{%
   \vcenter{\hbox{\rule[-.2pt]{#1}{.4pt}}}%
   \mkern-4mu\hbox{\usefont{U}{lasy}{m}{n}\symbol{41}}}}

\makeatletter

\setbox0\hbox{$\xdef\scriptratio{\strip@pt\dimexpr
    \numexpr(\sf@size*65536)/\f@size sp}$}

\newcommand{\scriptveryshortarrow}[1][3pt]{{%
    \vcenter{\hbox{\rule[\scriptratio\dimexpr-.2pt\relax]
               {\scriptratio\dimexpr#1\relax}{\scriptratio\dimexpr.4pt\relax}}}%
   \mkern-4mu\hbox{\let\f@size\sf@size\usefont{U}{lasy}{m}{n}\symbol{41}}}}

\title{Asymptotically Optimal Stochastic Lossy Coding of Markov Sources
\thanks{This work is supported in part by NSF under grant CCF-1909423.}
}

\author{
  Ahmed Elshafiy, Kenneth Rose \\
University of California, Santa Barbara \\
Santa Barbara, CA, 93117, USA \\
  \texttt{\{a\_elshafiy, rose\}@ece.ucsb.edu} \\
}

\begin{document}
\maketitle

\begin{abstract}
An effective ``on-the-fly'' mechanism for stochastic lossy coding of Markov sources using string matching techniques is proposed in this paper. Earlier work has shown that the rate-distortion bound can be asymptotically achieved by a ``natural type selection'' (NTS) mechanism which iteratively encodes asymptotically long source strings (from an unknown source distribution $P$) and regenerates the codebook according to a maximum likelihood distribution framework, after observing a set of $K$ codewords to ``$d$-match'' (i.e., satisfy the distortion constraint for) a respective set of $K$ source words. This result was later generalized for sources with memory under the assumption that the source words must contain a sequence of asymptotic-length vectors (or super-symbols) over the source super-alphabet, i.e., the source is considered a vector source. However, the earlier result suffers from a significant practical flaw, more specifically, it requires expanding the super-symbols (and correspondingly the super-alphabet) lengths to infinity in order to achieve the rate-distortion bound, even for finite memory sources, e.g., Markov sources. This implies that the complexity of the NTS iteration will explode beyond any practical capabilities, thus compromising the promise of the NTS algorithm in practical scenarios for sources with memory. This work describes a considerably more efficient and tractable mechanism to achieve asymptotically optimal performance given a prescribed memory constraint, within a practical framework tailored to Markov sources. More specifically, the algorithm finds asymptotically the optimal codebook reproduction distribution, within a constrained set of distributions having Markov property with a prescribed order, that achieves the minimum per letter coding rate while maintaining a specified distortion level.
\end{abstract}


\section{Introduction}
Stochastic codebook design, or equivalently quantizer design, based on source examples and string matching, has played a central role with different flavors in numerous applications in the areas of source coding, communications, machine learning, etc. Particularly influential were the seminal contributions of Lempel and Ziv in lossless coding, as evidenced by the numerous prevalent variants of the LZ77 and LZ78 algorithms \cite{LZ2,LZ3} which showed stochastic codebook generation/adaptation to be a powerful tool for lossless coding. Stochastic codebook generation mechanisms have been proposed for lossy coding as well, e.g., the gold-washing \cite{GW} and natural type selection \cite{NTS_original,NTS_parametric} algorithms. It is important to emphasize that optimizing the codebook reproduction distribution is fundamentally more difficult in the lossy coding setting. The lossless coding problem is ``simpler'' not only because perfect matching is less complex than matching with distortion, but also because the optimal codebook generating distribution, which achieves the minimum coding rate, is exactly the source distribution.  However, in lossy coding, the problem is vastly more difficult as the source distribution $P$ and optimal codebook generating distribution $Q^{*}$ are generally different, and more so in the high distortion regime \cite{LOSSY2,LOSSY3,NTS_original}. 

This idea of ``type selection'' for codebook generation or adaptation, which is the most relevant to the work done in this paper, was first introduced in the stochastic codebook generation and adaptation algorithm, known as ``Natural Type Selection'' (NTS), introduced in \cite{NTS_original,NTS_parametric} and further made practically tractable in \cite{NTS_ISIT,NTS_memory}. In this iterative algorithm, at each time step or iteration $n$, a sequence of $K$ independently generated source words, of length $L$, is encoded using a random codebook drawn from the generating distribution $Q_n$. For each source word in the sequence, the first codeword in the codebook to satisfy a specified distortion constraint $d$, is recorded. Then, the codebook reproduction distribution is updated for iteration $n+1$, by estimating the \textit{most likely distribution} to have generated the sequence of $K$ $d$-matching codewords. In other words, the codebook reproduction distributions (or types) are naturally selected in response to source examples, and evolve through a sequence of ``$d$-match'' operations, hence the name natural type selection, with a nod to Darwin's theory of evolution. Consequently, it was shown that asymptotically in, first, the statistical depth $K$, the number of iterations $n$, and the string length $L$, the sequence of codebook generating types $Q_1, Q_2, \dots$ converges to the optimal reproduction distribution $Q^{*}_{P,d}$ that achieves the rate-distortion bound $R(P,d)$ for memoryless sources. This result was further extended to sources with memory in \cite{NTS_memory}, by constructing source as a sequence of i.i.d. $M$-length \textit{vectors}, thus ignoring the inter-vector dependencies, for which the rate-distortion bound was achieved by a variant of the NTS algorithm, asymptotically as $M \rightarrow \infty$.

While ensuring asymptotic optimality, the NTS algorithm in \cite{NTS_memory} suffers from fundamental practical flaws. In order to converge to the optimal distribution that achieves the rate-distortion bound for sources with memory, the algorithm needs to encode source words that are composed of $M$-length vectors according to the $M$-th order source joint distribution $P_{M}$, while sending $M$ to infinity. Furthermore, even for finite-memory sources such as sources with finite order Markov property, the algorithm nevertheless requires sending the length of the i.i.d. source and code vectors $M$, within the $L$-length codeword, to infinity in order to achieve optimality. It is important to note that sending $M$ to infinity implies that the cardinalities of the source and code super alphabet spaces increase beyond any practical computational power available to perform $d$-search, ML estimation and codebook regeneration operations, thus rendering the system intractable in practical scenarios. The requirement of sending $M$ to infinity is also counter-intuitive, as it also applies to sources with finite memory, i.e., sources for which the current sample distribution only depends on a few of the past samples. In this paper, we propose to modify the NTS algorithm for Markov sources, such that the algorithm converges to its optimal distribution without sending $M$ to infinity. Specifically, we restrict the generating codebook distribution to $M$-th order Markov distributions, which may in practice be chosen to have the same order as the source. Then, asymptotic convergence to the optimal constrained distribution, i.e., the distribution that achieves the minimum per letter encoding rate amongst all codebook generating distributions of up to the prescribed Markov order, is shown.

The remainder of this paper is organized as follows: Section \ref{sec:background} covers the relevant background; Section \ref{sec:NTS_original} provides a review of the original NTS algorithm and highlights its limitations that motivate this work; Section \ref{sec:NTS_markov} discusses the practically-tractable and asymptotically optimal NTS algorithm tailored for Markov sources; Section \ref{sec:Toy_example} presents toy examples to illustrate the evolution of the codebook generating distribution for binary first order Markov source; Conclusions are drawn in Section \ref{sec:Conclusion}.

\section{Relevant Background on Random Coding \label{sec:background}}

Let $\{X_{u}\}_{u=1}^{\infty}$ be a stationary ergodic source, where the source realizations are denoted as $x_{u}\in\mathcal{X}$. We assume that the source alphabet $\mathcal{X}$, and the reproduction alphabet $\mathcal{Y}$ are discrete spaces, equipped with their associated Borel $\sigma$-field $\mathcal{X}'$, and $\mathcal{Y}'$, respectively. Furthermore, let $\{\mathbf{X}_{\ell}\}_{\ell=1}^{\infty}$ and $\{\mathbf{Y}_{\ell}\}_{\ell=1}^{\infty}$ be a sequence of i.i.d. $M$-tuples source and reproduction vectors,  where the realizations of source and reproductions vectors $\mathbf{{x}}_{\ell} \in \mathcal{X}^{M}$, and  $\mathbf{{y}}_{\ell} \in \mathcal{Y}^{M}$, respectively. Let $P_{M}$ denote the joint stationary distribution of source $M$-tuples $\mathbf{X}_{\ell}$. Define a random \textit{source word} $\mathbf{\tilde{X}}$ and a random \textit{codeword} $\mathbf{\tilde{Y}}$ as a concatenation of $L$ i.i.d. random source or reproduction vectors, respectively, i.e., $\mathbf{\tilde{X}} = [\mathbf{{X}}_{1}, \dots, \mathbf{{X}}_{L}]$, and $\mathbf{\tilde{Y}} = [\mathbf{{Y}}_{1}, \dots, \mathbf{{Y}}_{L}]$ Next, we define an arbitrary non-negative (measurable) scalar-valued distortion function $\rho\!:\! \mathcal{X} \times\!\!\ \mathcal{Y} \!\rightarrow\! [0,\infty)$. The distortion between a realization of the source word $\mathbf{\tilde{x}}$ and a realization of the codeword $\mathbf{\tilde{y}}$, is assumed additive, and is, specifically, the average distortion over samples:
\vspace{-0.5em}
\begin{equation}
    \rho\left(\mathbf{{x}}, \mathbf{{y}}\right) = \frac{1}{M} \sum\limits_{m=1}^{M} \rho\left({x}_{m}, {y}_{m}\right), \ \ \
    \rho\left(\mathbf{\tilde{x}},  \mathbf{\tilde{y}}\right) = \frac{1}{L} \sum\limits_{\ell=1}^{L} \rho\left(\mathbf{x}_{\ell},  \mathbf{y}_{\ell}\right).
    \label{eq:distortion_realization}
    \vspace{-0.5em}
\end{equation}
For a scalar-valued fidelity constraint $d$, define a ``${d}$-match" event as the event that $\rho\left(\mathbf{\tilde{x}}, \mathbf{\tilde{y}}\right) \leq d$. Suppose a random codebook $\mathcal{C}_{L}$ of infinite number of length-$ML$ codewords $\left(\mathbf{\tilde{Y}}(j), \text{ with } j\geq1 \right)$ is generated such that, each codeword consists of $L$ i.i.d. vectors as $Q_{M} = \{Q_{M}(\mathbf{y}): \mathbf{y} \in \mathcal{Y}^{M})\}$. We call $Q_{M}$ the codebook reproduction distribution. Let $N_{M,L}$ be the index of the first codeword in $\mathcal{C}_{L}$ that ${d}$-matches the source word realization $\mathbf{{x}}$, i.e., $N_{M,L} = \inf \left\{ j\geq1: \rho\left(\mathbf{\tilde{x}}, \mathbf{\tilde{y}}(j)\right) \leq d\right\}$,
with the convention that the infimum of an empty set is $+\infty$. Given a codebook reproduction distribution $Q_{M}$ over $\mathcal{Y}^{M}$, we define,
\begin{equation}
    D_{\mathrm{min}} \triangleq \mathbb{E}_{P_{M}}\left[\text{ess}\!\! \inf_{\mathbf{y}\sim Q_{M}} \rho(\mathbf{X},\mathbf{Y})\right], \ \ \ D_{\mathrm{av}} \triangleq \mathbb{E}_{P_{M}\times Q_{M}}\left[ \rho(\mathbf{X},\mathbf{Y})\right],
\end{equation}
where $\text{ess} \inf_{\mathbf{y}\sim Q_{M}}(\cdot)$ denotes the essential infimum of a function, i.e.,
\begin{equation}
    \text{ess}\!\! \inf_{\mathbf{Y}\sim Q_{M}} \rho(\mathbf{x},\mathbf{Y}) = \sup\{t \in \mathbb{R}: Q_{M}(\rho(\mathbf{x},\mathbf{Y})>t)=1 \}, \ \ \ \text{for any } \mathbf{x} \in \mathcal{X}^{M}.
\end{equation}
We will assume throughout this paper that $D_{\mathrm{av}}$ is finite, and $D_{\mathrm{min}}<D_{\mathrm{av}}<\infty$. We will also restrict our attention to the non-trivial range of distortion levels $d \in (D_{\mathrm{min}},D_{\mathrm{av}})$.
Then, Shannon's lossy coding theorem for scalar-valued distortion measures states: if a random codebook of length $\exp(L(R(P_{M},{d})+\epsilon))$ is generated using an optimal reproduction distribution $Q^{*}_{P_{M}, {d}}$, the probability of finding a codeword that ${d}$-matches an independently generated source word, drawn from the source distribution $P_{M}$, goes to one as $L$ goes to infinity, wherein $R(P_{M}, {d})$ is the \textit{joint} (or $M$-th order) rate-distortion function, i.e., \cite{Berger_RD,RD_Gray,Berger_RD_memory}
\begin{equation}
    R(P_{M},{d}) = \inf\limits_{\substack{V:[V]_{x} = P_{M}, \\   \mathbb{E}_{V}(\rho(\mathbf{X},\mathbf{Y}))\leq d}} I(\mathbf{X},\mathbf{Y}).
    \label{eq:RPD_cont}
\end{equation}
Here, $I(\mathbf{X},\mathbf{Y})$ is the mutual information between the $M$-tuples random vectors $\mathbf{X}$ and $\mathbf{Y}$, and the infimum is taken over all joint probability distributions $V$ such that the $x$-marginal of $V$, denoted $[V]_{x}$, is $P_{M}$ and the expected distortion $\mathbb{E}_{V}(\rho(\mathbf{X},\mathbf{Y})) \leq d$. Let $V^{*}_{P_{M},{d}}$ be the optimal joint distribution that realizes the infimum in (\ref{eq:RPD_cont}), then the optimal codebook reproduction distribution $Q^{*}_{P_{M},{d}}$ is the $y$-marginal of the optimal joint distribution $V^{*}_{P_{M},{d}}$. However, if a random codebook is generated from distribution $Q_{M} \neq Q^{*}_{P_{M},{d}}$, then the minimum encoding rate to guarantee a $d$-match in probability, as $L$ goes to infinity, was effectively shown in \cite{RPDQ}, and extended to memoryless sources over abstract alphabets in \cite{Zamir_NTS_cont}, to be
\begin{equation}
    \!R(P_{M},Q_{M},{d}) = \!\!\!\!\inf\limits_{\substack{V:[V]_{x} = P_{M}, \\   \mathbb{E}_{V}(\rho(\mathbf{X},\mathbf{Y}))\leq d}} \!\!\!\!  \mathcal{D}(V||P_{M} \times Q_{M}) = \inf_{Q_{M}'} \{ I_{\min}(P_{M}||Q_{M}', {d}) + \mathcal{D}(Q_{M}'||Q_{M}) \},\!\!
    \label{eq:RPQD_cont_orig}
\end{equation}
where $\mathcal{D}(\cdot||\cdot)$ is the Kullback-Leibler (KL) divergence (or the relative entropy), and $I_{\min}(P_{M}||Q_{M}', {d})$ is the usual minimum mutual information but with an additional constraint on the output distribution, i.e.,
\begin{equation}
    I_{\min}(P_{M}||Q_{M}', {d}) = \inf\limits_{\substack{V:[V]_{x} = P_{M}, \ [V]_{y} = Q_{M}', \\
    \mathbb{E}_{V}(\rho(\mathbf{X},\mathbf{Y}))\leq d}} I(\mathbf{X},\mathbf{Y}).
\end{equation}
Here the infimum is taken over all joint distributions $V$ of the random vectors $(\mathbf{X}, \mathbf{Y})$, whose $x$-marginal, denoted by $[V]_{x}$, is $P_{M}$, and $y$-marginal, denoted by $[V]_{y}$, is $Q_{M}'$, and such that the expected distortion does not exceed $d$. In \cite[Th. 2]{Kontoyiannis_string_matching}, it was shown that, under these assumptions for the memoryless case (for which extension to the sources with memory case is straight forward), $R(P_{M},Q_{M},{d})$ is finite, strictly positive, and
that the infimum in its L.H.S. definition in (\ref{eq:RPQD_cont_orig}) is always achieved by some joint distribution $V^*_{P_{M},Q_{M},{d}}$. Moreover, since the set of $V$ over which the infimum is taken is convex, from \cite{Csiszar_unique_minimizer} it can be concluded that $V^*_{P_{M},Q_{M},{d}}$ is the unique minimizer. Hence, a unique minimizer to the R.H.S. of (\ref{eq:RPQD_cont_orig}) also exists, and is denoted $Q^{*}_{P_{M},Q_{M},d}$.
Next, we define the minimum coding rate per letter for stationary ergodic sources with memory required to guarantee a ${d}$-match with probability one asymptotically in $L$ as \cite{RD_Gray} \cite{Berger_RD_memory}, 
\begin{equation}
    R({d}) = \lim\limits_{M \rightarrow \infty} M^{-1} R(P_{M},{d}).
    \label{eq:RDM}
\end{equation}
The limit in (\ref{eq:RDM}) exists for stationary ergodic sources, and for any $M$, $R(P_{M},{d})$ is an upper bound to $R({d})$ \cite[Th. 9.8.1]{Gallager_book}. Consequently, the optimal codebook reproduction distribution that achieves $R({d})$ is $Q^{*}_{{d}} = \lim\limits_{M\rightarrow \infty} Q^{*}_{P_{M}, {d}}$. Given a source with discrete input and reproduction alphabets, define a `\textit{type}' of source or code vector as the fraction of occurrence of every letter in the alphabet as seen in the vector \cite{Thomas_Cover}. To accommodate sources with memory, and account for memory depth of $M$, we proceed as follows \cite{NTS_memory}: define $Q_{n,M,L}\left(\mathbf{\tilde{y}}({j})\right) = \left\lbrace Q(\mathbf{y}): Q(\mathbf{y}) = \frac{1}{L} N(\mathbf{y}|\mathbf{\tilde{y}}({j})), \mathbf{y} \in {\mathcal{Y}}^{M}\right\rbrace$ as the \textit{$M$-th order type} of the $L$-length codeword $\mathbf{\tilde{y}}({j})$, where $N(\mathbf{y}|\mathbf{\tilde{y}}({j}))$ is the number of occurrences of the non-overlapping vector (or super symbol) $\mathbf{y}$ in the codeword. This is simply the type for the source considered as over a ``super alphabet'' of super-symbols.

In this paper, we restrict our attention to stationary and ergodic sources with memory described by the Markovian property. The $M$-th order Markov source property implies that the current source sample distribution, conditioned on the entire past, is fully captured by conditioning on the previous $M$-samples. This Markov source can be described by a state transition diagram, i.e., a Markov chain with exactly $|\mathcal{X}|^{M}$ states. Let $P_{j|i}$ be the homogenous source state transition probability from state $i$ to state $j$, where $i,j \in \mathcal{R} = \mathcal{X}^{M}$. Hence, let $\mathbf{P}$ be the state transition probability matrix for which the entry in the $i$-th row and $j$-th column is $P_{j|i}$. Equivalently, let $P(X|\mathbf{x}) = \{P(x|\mathbf{x}): x \in \mathcal{X}\}$ be the stationary source letter distribution conditioned on the $M$ previous samples specified in the vector $\mathbf{x}$. Note that there exists a one-to-one mapping between the set $\{P_{j|i}, \forall (i,j) \in \mathcal{R}^2 \}$ and the set $\{P(x|\mathbf{x}), \forall x \in \mathcal{X}, \forall \mathbf{x} \in \mathcal{X}^{M}\}$. By the stationary assumption of the Markov chain, the stationary marginal distribution is computed as,
\begin{equation}
   \mathbf{\Pi} = \mathbf{\Pi} \  \mathbf{P},
\end{equation}
where $\mathbf{\Pi} = [\pi(1), \dots, \pi(|\mathcal{R}|)]$, is a row vector representing the marginal distributions of all states in $\mathcal{R}$.

\section{Natural Type Selection \label{sec:NTS_original}}

In \cite{NTS_original}, a novel codebook regeneration algorithm was developed and shown to achieve asymptotically optimal performance, in the rate-distortion sense, for discrete memoryless sources. Consider the memoryless case for which $M$ is set to one, and the source letters are generated according to $P_{1} = \{P_{1}(x): x \in {\mathcal{X}}\}$, where ${\mathcal{X}}$ is a discrete-alphabet space. The subscript ``$1$'' here and below stands for $M=1$. Additionally, let the memoryless codebook reproduction distribution over discrete-alphabet space ${\mathcal{Y}}$, be $Q_{1} = \{Q_{1}(y): y \in {\mathcal{Y}}\}$. It was shown in \cite{NTS_original} that the empirical type of the codeword that $d$-matches an independently generated source word, converges in probability to $Q^{*}_{P_{1}, Q_{1}, d}$ as the string length $L\to\infty$. Note that $Q^{*}_{P_{1}, Q_{1}, d}$ is more efficient in coding the source than the initial $Q_{1}$, i.e., $R(P_{1}, Q^{*}_{P_{1},Q_{1},d},d) < R(P_{1}, Q_{1},d)$. This immediately suggests a recursive algorithm. Let $n$ be the iteration index, $ N_{1,L}$ be the index of the first $d$-matching codeword, whose type is denoted as $Q_{n,1,L}^{N_{1,L}}$. Starting with a strictly positive initial codebook reproduction distribution denoted $Q_{0,1,L}$, the type of the $d$-matching codeword at the current iteration is used to generate the codebook of the next iteration. In other words, the next iteration’s codebook reproduction distribution is naturally selected by the source through a $d$-match event, hence the name ``natural type selection''. The original NTS algorithm is summarized in Algorithm \ref{alg:NTS_original}. 
\begin{algorithm}
\caption{: Original NTS Algorithm}\label{alg:NTS_original}
\begin{algorithmic}[1]
\Procedure{NTS\_Original}{$N,L,d,Q_0,\mathbf{\tilde{x}}(1),\dots,\mathbf{\tilde{x}}(N)$}
\State $Q_{1,1,L}\gets Q_0.$
\For{$n=1:N$}
\State $i \gets n.$
\State $j\gets 0.$
\While{$d' < d$}
\State $j \gets j+1.$
\State Generate $j$-th codeword $\mathbf{\tilde{y}}(j)$ using $Q_{n,1,L}.$
\State $d'\gets \rho\left(\mathbf{\tilde{x}}(i),\mathbf{\tilde{y}}(j)\right).$
\EndWhile
\State $Q_{n+1,1,L}\gets Q_{n,1,L}^{N_{1,L}}.$
\EndFor
\State \textbf{return} $Q_{N+1,1,L}.$
\EndProcedure
\end{algorithmic}
\end{algorithm}

\noindent This algorithm results in a sequence of codebook reproduction distributions,
\begin{equation}
    Q_{n,1,L} = Q^{N_{1,L}}_{n-1,1,L}, 
    \vspace{-0.5em}
\end{equation}
\vspace{-0.5em}
\begin{equation}
    Q_{n,1} \!=\! \lim\limits_{L \rightarrow \infty}\! Q_{n,1,L} \!=\! Q^{*}_{P_{1}, Q_{n-1,1},d}, \ n = 1,2,\dots
    \label{eq:NTS_original}
\end{equation}
It was shown in \cite{NTS_original}, that after $L$ was taken to infinity, the sequence of distributions in (\ref{eq:NTS_original}), i.e., $Q_{0,1}, Q_{1,1}, Q_{2,1}, \dots$, coincides with the recursion in the fixed distortion version of the Arimoto and
Blahut (AB) algorithm \cite{Arimoto72,Blahut72} for computation of the rate-distortion function. Furthermore, \cite{NTS_original} provided a fixed-slope version of the NTS recursion, which coincide with the conventional form of Blahut algorithm \cite{Blahut72} \cite[Ch. 10]{Thomas_Cover}. Consequently, it was concluded that the NTS procedure stochastically simulates the AB algorithm, where the next distribution at each iteration step emerges “on the fly” through the coding process. It should be noted that Csiszar and Tusnady \cite{Alternating_Min} gave a useful geometric interpretation for the AB algorithms, as \textit{alternating projections} between convex sets, with the Kullback-Leibler distance replacing the usual Euclidean distance.

Hence, it was shown that the recursion in (\ref{eq:NTS_original}) converges to the optimal codebook distribution $Q^{*}_{P_{1}, d}$ that achieves the rate-distortion bound $R(P_{1}, d)$ in (\ref{eq:RPD_cont}) for $M=1$, i.e.,
\begin{equation}
    Q^{*}_{P_{1},d} = \lim\limits_{n\rightarrow \infty} \lim\limits_{L \rightarrow \infty} Q_{n,1,L},
\end{equation}
\begin{equation}
    R(P_{1}, d) = \lim\limits_{n \rightarrow \infty} \lim\limits_{L \rightarrow \infty} R(P_{1}, Q_{n,1,L}, d).
    \vspace{-0.3em}
\end{equation}

\noindent The asymptotic optimality result, established for the original NTS algorithm, suffers from several fundamental shortcomings impacting its practical implementation. The first shortcoming pertains to complexity and hinges on the order of limits that requires that string length $L$  be sent to infinity first, and only then can NTS iterations be performed ($n\to\infty$). In other words, NTS iterations must be performed on very large strings. Unfortunately, the probability of finding a $d$-match decreases exponentially with the string length, or alternatively, the codebook size must grow exponentially with $L$, which implies intractable $d$-search complexity, even in early NTS iterations. Clearly, it is the reversed order of limits that would be desirable in practice. In \cite{NTS_parametric}, a parametric set of codebook reproduction distributions $\mathcal{Q}_{\Theta}$ was considered, wherein the codebook reproduction distributions were constrained to a pre-specified family of distributions, $Q \in \mathcal{Q}_{\Theta}$, spanned by a parameter vector $\mathbf{\theta} \in \Theta$. A smoothing block on the parameter vector $\theta$ was proposed to reduce the fluctuations of the codebook reproduction distributions around the optimal solution, at finite string length $L$, which nevertheless exhibited some significant instability (see Figure 2 in \cite{NTS_parametric}). The above shortcomings represent a significant obstacle on the way to achieve major impact on lossy coding. The phenomenal impact of stochastic codebook generating algorithms (e.g., LZ78 in \cite{LZ3}), in many  lossless coding applications, suggests that overcoming these shortcomings may deliver considerable benefits. This provides a strong motivation to develop a practically tractable NTS algorithm that is asymptotically optimal for a wide spectrum of sources. 

In \cite{NTS_ISIT,NTS_memory}, we posed a natural and important questions: Can the convergence to the optimal reproduction distribution $Q^{*}_{P_{M},d}$ be achieved, but through a reversed order of limits? I.e., can we achieve $Q^{*}_{P_{M},d}$ by first sending $n$ to infinity, while maintaining finite $L$, and then sending $L$ to infinity? Obviously, if string length $L$ is finite, then the type of the $d$-matching codeword (of the same length) is restricted in resolution to $1/L$, as the relative frequency of a letter or super symbol, in the codeword, is a multiple of $1/L$. Such a low resolution of types may cause difficulties for an iterative algorithm that advances by potentially very small adjustments to the distribution. In order to circumvent this shortcoming, in \cite{NTS_ISIT}, we proposed to update the estimate of a general codebook reproduction distribution (not restricted to type resolution of $1/L$), after observing many $d$-matching events. In other words, we find the {\em maximum likelihood} estimate of the distribution that would have generated the observed sequence of $d$-matching codewords in response to a sequence of independently generated source words. Note that this approach is closely connected to finding the maximum likelihood codebook reproduction distribution, within a family of distributions, that maximize the probability of $d$-match events, which has been investigated in \cite{Kontoyannis_MLE1, Kontoyannis_MLE2}. However, in our method, we find the reproduction distribution that maximizes the probability of generating a set of observed $d$-matching codewords, which was shown to provide asymptotically optimal rate-distortion results as proven in \cite{NTS_ISIT,NTS_memory,NTS_abstract_alphabet}. Let $K$ be the number of the $d$-matching events considered before performing maximum likelihood estimation and updating the codebook reproduction distribution. Let the sequence of i.i.d. source vectors $\{\mathbf{x}_{\ell}\}_{\ell=1}^{\infty}$ be generated according to $P_{M} = \{P_{M}(\mathbf{x}): \mathbf{x} \in \mathcal{X}^{M}\}$. Furthermore, let the codebook reproduction distribution be $Q_{M} = \{Q_{M}(\mathbf{y}): \mathbf{y} \in \mathcal{Y}^{M}\}$. In \cite{NTS_memory}, we showed that the Maximum Likelihood (ML) distribution that most likely generates a sequence of $K$ codewords that respectively $d$-match a sequence of $K$ independently generated source words, converges in probability to $Q^{*}_{P_{M}, Q_{M}, d}$ as the statistical depth $K\to\infty$ and string length $L\to\infty$. Note that $Q^{*}_{P_{M}, Q_{M}, d}$ is more efficient in coding the source than $Q_{M}$. This immediately suggests a recursive and iterative algorithm. Let $n$ be the iteration index, and assume that the algorithm starts with a strictly positive initial codebook reproduction distribution denoted $Q_{0,M,L,K}$, over the entire reproduction space. At each iteration, the algorithm performance a sequence of $K$ $d$-match events to a sequence of $K$ independently generated source words. Afterwards, the algorithm computes the ML codebook distribution that would generate the set of $d$-matching codewords. In other words, the next iteration’s codebook reproduction distribution is naturally selected by the source through a sequence of $d$-match events, hence the name “natural type selection”. Denote $Q_{0,M,L,K}, Q_{1,M,L,K}, \dots$ as the sequence of ML codebook reproduction distributions, it was shown in \cite{NTS_memory} that this sequence of distributions converges to the optimal codebook distribution $Q^{*}_{P_{M}, d}$ that achieves the $M$-th order rate-distortion function $R(P_{M}, d)$ in (\ref{eq:RPD_cont}), i.e., $Q^{*}_{P_{M},d} = \lim\limits_{n\rightarrow \infty} \lim\limits_{L \rightarrow \infty} \lim\limits_{K \rightarrow \infty} Q_{n,M,L,K}$. The modified NTS algorithm is summarized in Algorithm \ref{alg:NTS_mod_disc}. In the next section, we introduce a variant of the NTS algorithm which is specialized for Markov sources and analyze the asymptotic optimality of its codebook generating distribution over all Markov distribution of the same order.

\begin{algorithm}
\caption{: Modified NTS Algorithm for Discrete Alphabets}\label{alg:NTS_mod_disc}
\begin{algorithmic}[1]
\Procedure{NTS\_Modified\_Discrete}{$N,M,L,K,d,Q_0,\mathbf{\tilde{x}}(1),\dots,\mathbf{\tilde{x}}(KN)$}
\State $Q_{1,M,L,K}\gets Q_0.$
\For{$n=1:N$}
\For{$k=1:K$}
\State $i \gets (n-1)K+k.$
\State $j\gets 0.$
\While{$d' < d$}
\State $j \gets j+1.$
\State Generate $j$-th codeword $\mathbf{\tilde{y}}(j)$ using $Q_{n,M,L,K}.$
\State $d'\gets \rho\left(\mathbf{\tilde{x}}(i),\mathbf{\tilde{y}}(j)\right).$
\EndWhile
\State Record $Q_k \gets M\text{-th order type of } \mathbf{\tilde{y}}(j).$
\EndFor
\State $Q_{n+1,M,L,K}\gets \frac{1}{K}\sum\limits_{k=1}^{K} Q_{k}.$
\EndFor
\State \textbf{return} $Q_{N+1,M,L,K}.$
\EndProcedure
\end{algorithmic}
\end{algorithm}

\section{Proposed NTS Algorithm for Markov Sources \label{sec:NTS_markov}}

In order to take into account the Markovian property of the source, we restrict the codebook reproduction distribution to distributions with $M$-th order Markov property. Let $Q_{j|i}$ be the codebook distribution state transition probability from state $i$ to state $j$, where $i \in \mathcal{S}, j \in \mathcal{S}, \text{ and } \mathcal{S} =  \mathcal{Y}^{M}$. Hence, let $\mathbf{Q}$ be the state transition probability matrix for which the entry in the $i$-th row and $j$-th column is $Q_{j|i}$. Let the random $L$-tuples source words and codewords $\mathbf{X} = [X_{1}, \dots, X_{L}]$, and $\mathbf{Y} = [Y_{1}, \dots, Y_{L}]$, be generated according to state transition matrices $\mathbf{P}$ and $\mathbf{Q}$, respectively. First, we introduce a variant of NTS algorithm for the above setup. At every NTS iteration with index $n$, the algorithm finds a set of $d$-matching codewords in the random codebook to a set of $K$ independently generated source words. Let the realizations of the $d$-matching source and code sets be denoted as $\{\mathbf{x}(i_1), \dots, \mathbf{x}(i_K)\}$, and $\{\mathbf{y}(j_{1}), \dots, \mathbf{y}(j_{K})\}$, where $j_{k}$ is the index of the codeword that $d$-match the $k$-th source word in the codebook. Next, similar to before, the NTS algorithm finds the most likely constrained reproduction distribution to produce the set of $d$-matching codewords, where the distribution is constrained to have $M$-th order Markov property. Note that in the codebook training stage, we assume that each source word (and hence each codeword) among the $K$-size set is generated independently of the other source words. This condition is necessary to guarantee convergence to the desired optimal constrained distribution, as will be illustrated in Theorem 1. However, once the training is completed and the codebook distribution has converged, this condition can be relaxed. Furthermore, for a given codebook distribution, the encoder and the decoder ``agree'' on a given random codebook by synchronizing the random number generator seed.

\begin{figure}
    \centering
    \includegraphics[scale=0.5]{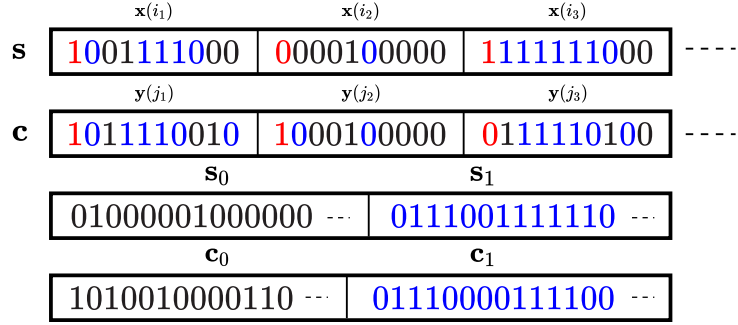}
    \caption{Division of the $d$-matching source and code blocks into i.i.d. sub-streams based on the previous sample.}
    \label{fig:markov_sub_streams}
\end{figure}

\textit{Lemma 1} \cite{Markov_MLE1}: The ML estimate of the $M$-th order Markov process state transition probabilities underlying the codebook reproduction distribution, given a set of $K$ $d$-matching codewords, is the \textit{average} of the $d$-matching codewords' transitions, i.e.,
\begin{equation}
    \mathbf{Q}_{n+1,M,L,K} = \mathbf{Q}^{\mathrm{ML}} = \left\lbrace Q_{j|i}:Q_{j|i}= \frac{\sum\limits_{k=1}^{K} N(i\!\veryshortarrow\! j|\mathbf{y}(k))}{\sum\limits_{k=1}^{K} \sum\limits_{j' \in \mathcal{S}} N(i \!\veryshortarrow\! j'|\mathbf{y}(k))}, \ \ \forall(i,j) \in \mathcal{S}^2 \right\rbrace,
    \label{eq:NTS_markov_recursion}
\end{equation}
where $k$ enumerates the $d$-matching events, and $N(j\!\veryshortarrow\!i|\mathbf{y}(j_k))$ is the number of transitions from state $i$ to state $j$ as seen in the $k$-th $d$-matching codeword ${{\mathbf{{y}}} ({j_k})}$, whose index in the random codebook is $j_k$. The detailed proof of Lemma 1 is provided in Appendix A. Thus, this algorithm yields a sequence of state transition matrices as in (\ref{eq:NTS_markov_recursion}), or equivalently, a sequence of conditional distributions $Q_{n+1,M,L,K}(Y|\mathbf{y})$. In the next discussion, we quantify the asymptotic performance of the NTS algorithm specialized for Markov sources. Let the random codebook be generated according to a Markov process with conditional probabilities $Q(Y|\mathbf{y}), \ \forall \mathbf{y} \in \mathcal{Y}^{M}$, i.e., $Q(Y|\mathbf{y})$ is the row in the matrix $\mathbf{Q}$ that corresponds to transitions from state $s = \mathbf{y} \in \mathcal{S}$. We start by transforming this variant of NTS algorithm into a dual set of NTS algorithms for \textit{memoryless sources}. Let the sets of $K$ $d$-matching source words and codewords be concatenated into $KL$-length source and code blocks denoted as, $\mathbf{s} = [\mathbf{x}(i_1), \dots, \mathbf{x}(i_K)]$, and $\mathbf{c} = [\mathbf{y}(j_1), \dots, \mathbf{y}(j_K)]$, respectively. Furthermore, let the source and code blocks be independently divided into sub-streams based on the previous source and code $M$-tuples, denoted as $\{\mathbf{s}_\mathbf{x}, \forall \mathbf{x} \in \mathcal{X}^{M}\}$, and $\{\mathbf{c}_\mathbf{y}, \forall \mathbf{y} \in \mathcal{Y}^{M}\}$. Hence, the first $M$ letters in each of the source and code words are not assigned to any sub-stream, which asymptotically in $L$ have negligible effect. Note that by the homogenous assumption of the Markovian source and code distributions, the sequence of symbols of the sub-streams $\{\mathbf{s}_\mathbf{x}\}$ are i.i.d. according to $P(X|\mathbf{x})$. The set of $d$-match event $\{\rho(\mathbf{{x}}(i_k), \mathbf{{y}}(i_k)) \leq d, \ \ \forall k\}$, implies that $\rho(\mathbf{s}, {\mathbf{c}})\leq d$, which is equivalent to a set of size $|\mathcal{X}^{M} \times \mathcal{Y}^{M}|$ events of distortion matches between the sub-streams $\{\mathbf{s}_{\mathbf{x}}\}$ and $\{\mathbf{c}_{\mathbf{y}}\}$, each with distortion level denoted as $d_{\mathbf{x},\mathbf{y}}$, such that, $\sum_{\mathbf{x},\mathbf{y}} \mathbb{M}_{n}(\mathbf{x},\mathbf{y})d_{\mathbf{x},\mathbf{y}} \leq d$. Here $\mathbb{M}_{n}(\mathbf{x},\mathbf{y})$, is the empirical probability of mapping a letter in sub stream $\mathbf{{s}}_{\mathbf{x}}$, to a letter in sub stream $\mathbf{{c}}_{\mathbf{y}}$, at NTS iteration $n$, as seen by the code and source blocks $\mathbf{s}$, and $\mathbf{c}$, respectively. To illustrate this further, consider Fig \ref{fig:markov_sub_streams} showing binary source and code blocks, which are formed by concatenating three $d$-matching $L$-length source and code words, with $L=10$, where the source and code generating distributions exhibit a first order Markovian property, hence the number of Markov states is $|\mathcal{X}| = |\mathcal{Y}|= 2$. Furthermore, a Hamming distortion measure is employed at distortion level $d=1/3$. Samples in different i.i.d. sub-streams are assigned different colors. Samples whose immediate predecessor is a  `$0$' are colored black, and samples following a `$1$' are colored blue. The i.i.d sub-streams $\mathbf{s}_{\mathbf{x}}$ and $\mathbf{c}_{\mathbf{y}}$ are formed by collecting all samples that follow the same letter, as shown in Fig. \ref{fig:markov_sub_streams}. Hence, the iterative NTS algorithm for Markovian sources over discrete alphabet spaces can be summarized in Algorithm \ref{alg:NTS_mod_markov_disc}.

\begin{algorithm}
\caption{: Modified NTS Algorithm for Markovian Sources over Discrete Alphabets}\label{alg:NTS_mod_markov_disc}
\begin{algorithmic}[1]
\Procedure{NTS\_Markov\_Discrete}{$N,M,L,K,d,\mathbf{Q}_0,\mathbf{{x}}(1),\dots,\mathbf{{x}}(KN)$}
\State $\mathbf{Q}_{1,M,L,K}\gets \mathbf{Q}_0.$
\For{$n=1:N$}
\For{$k=1:K$}
\State $i \gets (n-1)K+k.$
\State $j\gets 0.$
\While{$d' < d$}
\State $j \gets j+1.$
\State Generate $j$-th codeword $\mathbf{y}(j)$ using state transition matrix $\mathbf{Q}_{n,M,L,K}$.
\State $d'\gets \rho\left(\mathbf{{x}}(i),\mathbf{{y}}(j)\right).$
\EndWhile
\State Record $N^{(i\scriptveryshortarrow j)}_{k} \gets N(i \!\veryshortarrow \! j|\mathbf{y}(j)), \ \ \ \forall (i, j) \in \mathcal{S}^2.$
\EndFor
\State $\mathbf{Q}_{n+1,M,L,K} \gets \left\lbrace Q_{j|i}:Q_{j|i}= \frac{\sum\limits_{k=1}^{K} N^{(i \scriptveryshortarrow j)}_{k}}{\sum\limits_{k=1}^{K} \sum\limits_{j' \in \mathcal{S}} N^{(i\scriptveryshortarrow j')}_{k}}, \ \ \forall (i,j) \in \mathcal{S}^2 \right\rbrace.$
\EndFor
\State \textbf{return} $\mathbf{Q}_{N+1,M,L,K}.$
\EndProcedure
\end{algorithmic}
\end{algorithm}

 \textbf{Theorem 1:} For an initial codebook generating Markov chain with strictly positive transition probabilities $Q(Y|\mathbf{y})>0$,  $\forall \mathbf{y} \in \mathcal{S} = \mathcal{Y}^{M}$, and distortion measure satisfying $0\leq D_{\mathrm{min}}<D_{\mathrm{av}}<\infty$, the transition probabilities ${Q}(Y|\mathbf{y})$, of the recursive NTS algorithm for Markov sources, where each recursion involves collecting $K$ $d$-matches, converge in probability and asymptotically, as $L \rightarrow \infty$, as follows,
\begin{equation}
\begin{aligned}
    &{Q}_{n+1,M,K}(Y|\mathbf{y}) \!\rightarrow \!\!\!\sum\limits_{\mathbf{x} \in \mathcal{X}^{M}}\!\! {\mathbb{M}}^{*}_{n}(\mathbf{x}|\mathbf{y}) Q^{*}\left(P(X|\mathbf{x}),{Q}_{n,M,K}(Y|\mathbf{y}),{d}^{*}_{\mathbf{x},\mathbf{y}}\right)\!,\!\! \\[-5pt]
      &V^{*}\left(P(X|\mathbf{x}),{Q}(Y|\mathbf{y}),{d}^{*}_{\mathbf{x},\mathbf{y}}\right) \triangleq \arg \min\limits_{V \in E_{\mathbf{x},\mathbf{y}}({d}^{*}_{\mathbf{x},\mathbf{y}})} \!\!\!\!\mathcal{D}\left(V \Big| \Big| P(X|\mathbf{x}) \!\times\!{Q}(Y|\mathbf{y})\right), \\[-5pt]
     & Q^{*}\left(P(X|\mathbf{x}),{Q}(Y|\mathbf{y}),{d}^{*}_{\mathbf{x},\mathbf{y}}\right) = \left[ V^{*}\left( P(X|\mathbf{x}),{Q}(Y|\mathbf{y}),{d}^{*}_{\mathbf{x},\mathbf{y}}\right)\right]_{y},
    \label{eq: NTS_modified_4}
\end{aligned}
\end{equation}
where ${Q}_{n+1,M,K}(Y|\mathbf{y}) = \lim\limits_{L\rightarrow\infty} {Q}_{n+1,M,L,K}(Y|\mathbf{y})$, and the set $E_{\mathbf{x},\mathbf{y}}({d}_{\mathbf{x},\mathbf{y}})$ is defined as,
\begin{equation}
    E_{\mathbf{x},\mathbf{y}}({d}^{*}_{\mathbf{x},\mathbf{y}}) = \left\lbrace V: V = P' \circ W', P' = {P}(X|\mathbf{x}), \ \rho(P',W') \leq {d}^{*}_{\mathbf{x},\mathbf{y}} \right\rbrace.
    \label{eq:E_x_y}
\end{equation}
Here $\rho(P',W')$ is the average distortion computed over distributions, and the set of distortion levels $\{{d}^{*}_{\mathbf{x},\mathbf{y}}, \forall \mathbf{x}\in\mathcal{X}^{M}, \forall \mathbf{y} \in \mathcal{Y}^{M} \}$, satisfies,
\begin{equation}
    \frac{\partial}{\partial \delta} R'(P(X|\mathbf{x}),{Q}(Y|\mathbf{y}), \delta{d}^{*}_{\mathbf{x},\mathbf{y}})\Big|_{\delta=
{d}^{*}_{\mathbf{x},\mathbf{y}}} \!\!\!\!\!= R'_{P,Q,d}, \ \forall (\mathbf{x},\mathbf{y}),\quad  \sum\limits_{\mathbf{x},\mathbf{y}} {\mathbb{M}}^{*}_{n}(\mathbf{x},\mathbf{y}) {d}^{*}_{\mathbf{x},\mathbf{y}} \leq d,
    \label{eq:equal_slopes}
\end{equation}
where $R'_{P,Q,d}$ is independent of the sub-stream pair $(\mathbf{x},\mathbf{y})$. In other words, the distortion allocation to sub-stream pairs, $d^{*}_{\mathbf{x},\mathbf{y}}$, ensures they all maintain the same rate-distortion slope, given codebook generating distributions $\{Q(Y|\mathbf{y})\}$, while satisfying the overall average distortion constraint $d$. A brief proof sketch is that, we employ a variant of the conditional limit theorem in \cite{Thomas_Cover} to establish that, conditioned on the rare event that the joint input-output distribution of a block of $K$ concatenated respective source and codewords $\left(\mathbf{S}, \mathbf{C}\right)$ belongs to a convex set of distributions that satisfy the distortion constraint $d$, then the conditional distributions of this code block on $\mathcal{Y}$ converge in probability, as $L \rightarrow \infty$, to the distribution $\sum\limits_{\mathbf{x} \in \mathcal{X}^{M}}\!\! {\mathbb{M}}^{*}_{n}(\mathbf{x}|\mathbf{y}) Q^{*}\left(P(X|\mathbf{x}),{Q}_{n,M,K}(Y|\mathbf{y}),{d}^{*}_{\mathbf{x},\mathbf{y}}\right)$. Next, by \cite{RD_Gray}, the minimum of the output-constrained rate is achieved by adding the output-constrained rate-distortion functions at points of equal slopes in all co-ordinates, implying (\ref{eq:equal_slopes}).

This theorem has an intimate relationship with the \textit{Gibbs Conditioning Principle} of statistical mechanics (see \cite{Gibbs_conditioning_principle} and the references therein). The Gibbs conditioning principle roughly states: suppose that $\{X_{1} ,\dots, X_{N}\}$ are i.i.d. random variables distributed over a Polish space with marginal distribution $P_{X}$ and a measurable function $f:\mathcal{X} \rightarrow \mathbb{R}$. Hence, under suitable conditions on $P_{X}$ and $f(\cdot)$, and conditioned on the rare event that $\left\lbrace\frac{1}{N}\sum\limits_{i} f(X_{i}) \in [a-\delta, a+\delta]\right\rbrace$, where $a \in \mathbb{R}$ and $\delta>0$, the distribution of $X_{i}$ converges in probability, as $N\rightarrow\infty$, to the distribution that minimizes the divergence $\mathcal{D}(\cdot||P_{X})$ over all distributions that satisfy the constraint, which is very closely related to the arguments used to prove Theorem 1. The detailed proof of Theorem 1 is provided in Appendix B. Next, we look at the asymptotic convergence of the codebook reproduction conditional distributions as the number of iterations $n$ goes to infinity.

\textbf{Theorem 2:} Given an initial codebook that is generated using a Markov process with strictly positive conditional distributions $Q(Y|\mathbf{y})$ for any state $\mathbf{y} \in \mathcal{S} = \mathcal{Y}^{M}$, the recursion in (\ref{eq:NTS_markov_recursion}) achieves the minimum average coding rate over the cross product of all source-code sub streams, denoted as $\overline{R}(d)$, i.e.,
\begin{equation}
    \overline{R}(d)= \min\limits_{Q(Y|\mathbf{y})} \min\limits_{\substack{\mathbb{{M}}(\mathbf{y}|\mathbf{x})\\
    {d}_{\mathbf{x},\mathbf{y}}, V_{\mathbf{x},\mathbf{y}}}} \sum\limits_{\mathbf{x},\mathbf{y}} \mathbb{M}(\mathbf{x}) \mathbb{M}(\mathbf{y}|\mathbf{x}) \ \mathcal{D}\left(V_{\mathbf{x},\mathbf{y}} \ \big|\big| \  P(X|\mathbf{x}) \times Q(Y|\mathbf{y})\right).
    \label{eq:avg_rate_markov}
    \vspace{-0.5em}
\end{equation}
and the set of optimization variables that achieves the minimum in (\ref{eq:avg_rate_markov}), satisfies,
\vspace{-0.8em}
\begin{equation}
    \!\frac{\partial}{\partial \delta}R\left(P(X|\mathbf{x}), Q^{*}(Y|\mathbf{y}), \delta\right)\!\Big|_{\delta = {d}^{*}_{\mathbf{x}, \mathbf{y}}} \!\!\!\!= R'_{P,Q^{*},d}, \ \sum\limits_{\mathbf{x}, \mathbf{y}} \mathbb{{M}}^{*}(\mathbf{x},\mathbf{y}) {d}^{*}_{\mathbf{x}, \mathbf{y}} \leq d ,\!   \label{eq:theorem_2}
\end{equation}
where $R'_{P,Q^{*},d}$ is independent of the sub-stream pair $(\mathbf{x},\mathbf{y})$.
The proof sketch of Theorem 2 is, first, we show that the average encoding rate over the cross product of all i.i.d. source and code sub-streams can be written as double minimization over \textit{convex} sets, due to the convexity of all constraints. Then, we invoke the Csiszar and Tusnady theorem of alternating minimization over convex sets in \cite{Alternating_Min} to show the convergence of the average encoding rate to its minimum, and consequently, the convergence of the conditional distributions to distributions that achieves the minimum in \ref{eq:avg_rate_markov}. Details of Theorem's 2 proof is shown in Appendix C. Hence, this establishes that the NTS algorithm finds the conditional distributions that minimize the average encoding rate the cross product of over all i.i.d. source and code sub-streams $\{\mathbf{s}_{\mathbf{x}} \times \mathbf{c}_{\mathbf{y}}\}$ while maintaining the distortion level $d$, hence implying asymptotic optimality.

\section{Toy Example \label{sec:Toy_example}}

\begin{figure}[tb]
    \centering
    \includegraphics[scale=0.38]{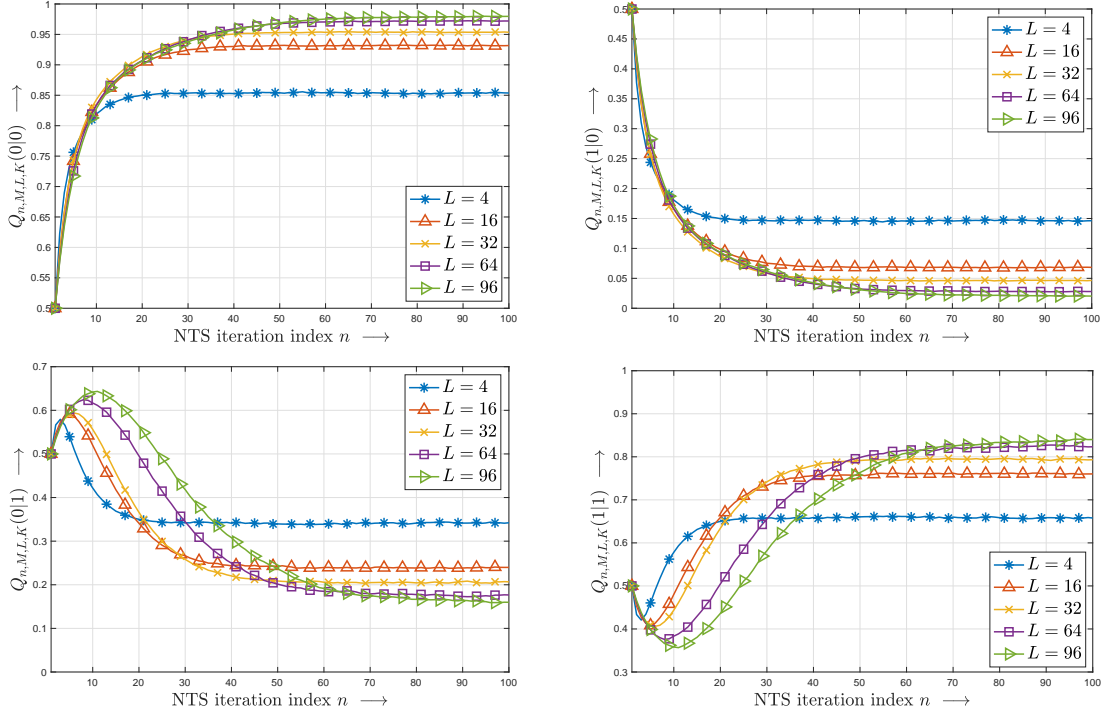}
    \vspace{-1.5em}
    \caption{Evolution of state transition probabilities in the NTS codebook generating Markov chain versus iteration index $n$, for different source word lengths $L$, statistical depth is $K=10^5$; operating on a binary Markov source with Hamming distortion constraint at $d = d_{\max} = 1/3$.}
    
    \label{fig:toy_example_memory_markov}
    \vspace{-0.5em}
\end{figure}

In this section, we illustrate the convergence behavior of the proposed NTS algorithm variant, which is tailored to Markovian sources. We consider a first-order binary Markov source with the transition probabilities $P(0|0) = 0.8,  P(1|0) = 0.2, P(0|1) = 0.4$, and $P(1|1) = 0.6$, to be encoded under the Hamming distortion measure. To illustrate how the algorithm operates, we choose a distortion constraint for which one can guess the optimal solution, $d = d_{\max} = 1/3$. In Fig. \ref{fig:toy_example_memory_markov}, we depict the evolution of transition probabilities of the codebook reproduction distributions versus the number of NTS iterations $n$, for different values of source word length $L$ . It is worth noting that as $L$ increases, and for the given distortion level $d=d_{\max}$, the transition probabilities of the codebook reproduction distribution approaches $Q^{*}(0|0) = 1$, which strongly favors the optimal codeword achieving $d_{\max}$, namely the all zero codeword. Thus, NTS is converging to the optimal first-order Markov codebook generating distribution. Furthermore, it is important to note that, even for finite length $L$, the codebook generating distributions converge asymptotically in the statistical depth $K$, and the number of NTS iterations $n$.

\section{Conclusion \label{sec:Conclusion}}

This paper proposes a modified and more effective NTS approach for a stochastic generation of random codebook in the lossy coding settings, specifically when sources with Markovian property are considered. Unlike the NTS approach in \cite{NTS_memory}, the algorithm is not required to send the memory depth $M$ to infinity in order to achieve the rate-distortion bound, which consequently reduces dramatically the complexity of finding a $d$-match in the codebook, the central operation of the NTS algorithm. It was further shown by Theorem 1 and Theorem 2, that the proposed codebook generating distribution, that emerges from the proposed stochastic algorithm, converges to the optimal codebook reproduction distribution of the prescribed order, asymptotically as $L \rightarrow \infty$, $K \rightarrow \infty$, and $n \rightarrow \infty$. 

\appendix

\section*{Appendix A -
Proof of Lemma 1: ML Estimation of Codebook  Distribution \label{app_lemma}}

The maximum likelihood estimator for the codebook distribution with Markov property is formulated as,

\begin{equation}
    \mathbf{Q}^{\mathrm{ML}} = \arg \max_{\mathbf{Q} \in \mathcal{Q}} \mathbb{P}(\mathbf{y}(j_1), \dots, \mathbf{y}(j_k)|\mathbf{Q}),
\end{equation}
where $\mathcal{Q}$ is the set of all valid Markov transition matrices that satisfy the stationary assumption. Next, by the independence of the $d$-match events, the maximum likelihood formulation is written as,

\begin{equation}
  \mathbf{Q}^{\mathrm{ML}} = \arg \max_{\mathbf{Q} \in \mathcal{Q}} \prod\limits_{k=1}^{K}\mathbb{P}(\mathbf{y}(j_k)|\mathbf{Q}).
\end{equation}
Next, we have,
\begin{equation}
    \mathbb{P}(\mathbf{y}|\mathbf{Q}) = \prod\limits_{m=1}^{M}\mathbb{P}(Y_{m} = y_{m} \ \! | \ \!  Y_{m-1} = y_{m-1}, \dots, Y_{1} = y_{1}) \prod\limits_{i,j \in \mathcal{S}} Q_{j|i}^{N(i \ \! \scriptveryshortarrow j|\mathbf{y})}.
\end{equation}
\noindent
Ignoring the first product term for sufficiently long codeword lengths $L$, and taking the log of $\mathbb{}{P}(\mathbf{y}|\mathbf{Q})$ results in,
\begin{equation}
  \mathbf{Q}^{\mathrm{ML}} \approx \arg \max_{\mathbf{Q} \in \mathcal{Q}} \left\lbrace  \sum\limits_{k=1}^{K} \sum\limits_{i,j \in \mathcal{S}} N(i \!\veryshortarrow\! j|\mathbf{y}(k)) \log(Q_{j|i}) \right\rbrace.
\end{equation}
The Lagrangian function that enforces the set of constraints, $\sum\limits_{j \in \mathcal{S}} Q_{j|i} = 1$, can be written as,
\begin{equation}
    \mathcal{L} = \sum\limits_{k=1}^{K} \sum\limits_{i,j \in \mathcal{S}} N(i\!\veryshortarrow\!j|\mathbf{y}(k))\log(Q_{j|i}) - \sum_{i \in \mathcal{S}} \lambda_{i} \left( \sum\limits_{j \in \mathcal{S}} Q_{j|i} - 1\right).
\end{equation}
Finally, Lemma 1 follows by differentiating with respect to $Q_{j|i}$, and setting to zero.

\section*{Appendix B -
Conditional Limit Theorem for NTS Algorithm with Markov Sources \label{App_theorem_1}}
In the subsequent analysis, without loss of generality, we will only consider the case for which $K=1$. Note that the distribution of the long source and code blocks $\mathbf{s}$ and ${\mathbf{c}}$ (formed by concatenating $K$ $d$-matching source and code words, respectively) do not change for $K>1$, as $L \rightarrow \infty$, because it is the $K$-average of identical converging distributions, as will be shown by this theorem. Let the length $L_{\mathbf{x},\mathbf{y}}$ be the number of letters of the source sub-stream $\mathbf{s}_{\mathbf{x}}$ that is reproduced by the code sub-stream $\mathbf{c}_{\mathbf{y}}$, such that $\sum L_{\mathbf{x},\mathbf{y}} = L$. For asymptotically large $L_{\mathbf{x},\mathbf{y}}$, we define the set $E_{{\mathbf{x},\mathbf{y}}}(d_{\mathbf{x},\mathbf{y}})$ as in (\ref{eq:E_x_y}). Thus $E_{{\mathbf{x},\mathbf{y}}}(d_{\mathbf{x},\mathbf{y}})$ denotes a set of all joint distributions on $\mathcal{X} \times \mathcal{Y}$ that satisfies the source distribution $P(X|\mathbf{x})$ and distortion level $d_{\mathbf{x},\mathbf{y}}$. Note that, by the strong law of large numbers, the realizations of the instantaneous source types $P_{L_{\mathbf{x},\mathbf{y}}}$ of the source letters in the sub-stream $\mathbf{s}_{\mathbf{x}}$ that were represented by code letters in the sub-stream $\mathbf{c}_{\mathbf{y}}$, converge almost surely to $P(X|\mathbf{x})$. Next, define the minimum divergence $\mathcal{D}^*$ as,
\begin{equation}
    \mathcal{D}^{*} = \min\limits_{\substack{\mathbb{{M}}(\mathbf{y}|\mathbf{x})\\
    {d}_{\mathbf{x},\mathbf{y}}, V_{\mathbf{x},\mathbf{y}}}} \sum\limits_{\mathbf{x} \in \mathcal{X}^{M}} \mathbb{M}(\mathbf{x}) \sum\limits_{\mathbf{y} \in \mathcal{Y}^{M}} \mathbb{{M}}(\mathbf{y}|\mathbf{x}) \ \mathcal{D}(V_{\mathbf{x},\mathbf{y}}||P(X|\mathbf{x}) \times Q(Y|\mathbf{y})), 
    \label{eq:min_average_div}
\end{equation}
such that $\mathbb{M}(\mathbf{y}|\mathbf{x})$, $d_{\mathbf{x},\mathbf{y}}$ and $V_{\mathbf{x},\mathbf{y}}$ satisfy,
\begin{equation}
    \sum\limits_{\mathbf{x},\mathbf{y}} \mathbb{M}(\mathbf{x})  \mathbb{M}(\mathbf{y}|\mathbf{x}) \rho(x_{1},y_{1}) = \sum\limits_{\mathbf{x} ,\mathbf{y}} \mathbb{M}(\mathbf{x})  \mathbb{M}(\mathbf{y}|\mathbf{x}) d_{\mathbf{x},\mathbf{y}} \leq d, \  V_{\mathbf{x},\mathbf{y}} \in E_{\mathbf{x},\mathbf{y}}({d_{\mathbf{x},\mathbf{y}}}),
    \label{eq:conditions_1}
\end{equation}
with $x_{1}$ and $y_{1}$ being the left most letters in $\mathbf{x}$ and $\mathbf{y}$, respectively. Let $\mathbb{M}^{*}(\mathbf{y}|\mathbf{x})$, $d^{*}_{\mathbf{x},\mathbf{y}}$, and $V^{*}_{\mathbf{x},\mathbf{y}}$ be the optimization variables that achieve the minimum in (\ref{eq:min_average_div}). Furthermore, define $\mathcal{D}^{*}_{\mathbf{y}} \triangleq \sum\limits_{\mathbf{x}} \mathbb{M}^{*}(\mathbf{x}|\mathbf{y}) \mathcal{D}(V^{*}_{\mathbf{x},\mathbf{y}}||P(X|\mathbf{x}) \times Q(Y|\mathbf{y})), \forall \mathbf{y} \in \mathcal{Y}^{M}$,
where $\mathbb{M}^{*}(\mathbf{x}|\mathbf{y})$ can be calculated from $\mathbb{M}(\mathbf{x})$ and $\mathbb{M}^{*}(\mathbf{y}|\mathbf{x})$ from Bayes' law. Hence, we have, $\mathcal{D}^{*} = \sum\limits_{\mathbf{y}} \mathbb{M}(\mathbf{y})\mathcal{D}^{*}_{\mathbf{y}}$, with $\mathbb{M}(\mathbf{y}) = \sum\limits_{\mathbf{x}'} \mathbb{M}(\mathbf{x}')\mathbb{M}^{*}(\mathbf{y}|\mathbf{x}')$. We will show that for any $\delta > 0$, and sufficiently large $L = \sum\limits_{\mathbf{x},\mathbf{y}} L_{\mathbf{x},\mathbf{y}}$,
\begin{equation}
    \begin{aligned}
    &\mathbb{P}\left(\sum\limits_{\mathbf{x}}\mathbb{M}(\mathbf{x}|\mathbf{y})\mathcal{D}(Q_{\mathbf{{C}}|\mathbf{x},\mathbf{y}}|| Q^{*}(P(X|\mathbf{x}), Q(Y|\mathbf{y}), {d}^{*}_{\mathbf{x},\mathbf{y}})) > 3\delta | V_{\mathbf{S},\mathbf{C}} \!\in\! E_{L}(d)\right) \\[-5pt]
    & \ \ \ \ \ \ \ \ \ \ \ \ \ \ \ \ \ \ \ \ \ \ \ \ \ \ \ \ \ \ \ \  \ \ \ \ \ \ \ \ \ \ \ \leq \prod\limits_{\mathbf{x}}(L\mathbb{M}(\mathbf{y})\mathbb{M}^{*}(\mathbf{x}|\mathbf{y}) + 1)^{2|\mathcal{X}||\mathcal{Y}|}e^{-L\delta}.
    \end{aligned}
\end{equation}
In other words, if we condition on the event that the joint distribution of the random source and code block pair (with $K=1$) $\mathbf{{S}}$, and $\mathbf{{C}}$, denoted as $V_{{\mathbf{S}},{\mathbf{C}}}$ over $\mathcal{X} \times \mathcal{Y}$, belongs to $E_{L}(d)$, the average conditional distribution of the codewords is with high probability close, in the divergence-sense, to $\sum\limits_{\mathbf{x}}\mathbb{M}(\mathbf{x}|\mathbf{y}) Q^{*}(P(X|\mathbf{x}), Q(Y|\mathbf{y}), {d}^{*}_{\mathbf{x},\mathbf{y}})$. Since closeness in divergence also implies closeness in the $\mathcal{L}_{1}$ sense \cite[Lemma 11.6.1]{Thomas_Cover}, this establishes Theorem 1. Then following \cite[Th. 11.6.2]{NTS_ISIT,Thomas_Cover}, we obtain,
\begin{equation}
    \begin{aligned}
    &\mathbb{P}\left(\sum\limits_{\mathbf{x}}\mathbb{M}(\mathbf{x}|\mathbf{y})\mathcal{D}\left(V_{\mathbf{{S}},\mathbf{{C}}|\mathbf{x},\mathbf{y}}|| P(X|\mathbf{x}) \!\times Q(Y|\mathbf{y})\right) \!>\! D^{*} \!+ 3\delta, V_{\mathbf{{S}},\mathbf{{C}}} \in E_{L}(d)\right) = \\[-6pt]
    & \ \ \ \ \ \ \sum\limits_{\substack{\\[-5pt] V^{'}_{\mathbf{x},\mathbf{y}} \in E_{{\mathbf{x},\mathbf{y}}}({d}_{\mathbf{x},\mathbf{y}}) \cap \mathcal{P}_{L_{\mathbf{x},\mathbf{y}}} \times \mathcal{Q}_{L_{\mathbf{x},\mathbf{y}}}: \\ \sum\limits_{\mathbf{x}}\mathbb{M}(\mathbf{x}|\mathbf{y})\mathcal{D}(V^{'}_{\mathbf{x},\mathbf{y}}||P(X|\mathbf{x})\times Q(Y|\mathbf{y}))>D^{*}_{\mathbf{y}} + 3\delta \\
    \sum\limits_{\mathbf{x},\mathbf{y}}\mathbb{M}(\mathbf{x},\mathbf{y}) \rho(x_{1},y_{1}) = \sum\limits_{\mathbf{x},\mathbf{y}}\mathbb{M}(\mathbf{x},\mathbf{y}) d_{\mathbf{x},\mathbf{y}}\leq d}} \prod\limits_{\mathbf{x}}\mathbb{P}(T_{\mathrm{y}}({V^{'}_{\mathbf{x},\mathbf{y}}})),
    \end{aligned}
\end{equation}
where the probability of type class of $V^{'}_{\mathbf{x},\mathbf{y}}$ is denoted by $\mathbb{P}(T_{\mathrm{y}}({V^{'}_{\mathbf{x},\mathbf{y}}}))$, $\mathcal{P}_{L_{\mathbf{x},\mathbf{y}}}$, and $\mathcal{Q}_{L_{\mathbf{x},\mathbf{y}}}$ are the sets of all possible input and output types with denominator $L_{\mathbf{x},\mathbf{y}}$. Then, by  \cite[Th. 11.1.4]{Thomas_Cover} which bounds the probability of type classes,
\begin{equation}
    \begin{aligned}
    &\mathbb{P}\left(\sum\limits_{\mathbf{x}}\mathbb{M}(\mathbf{x}|\mathbf{y})\mathcal{D}\left(V_{\mathbf{{S}},\mathbf{{C}}|\mathbf{x},\mathbf{y}}|| P(X|\mathbf{x}) \!\times Q(Y|\mathbf{y})\right) \!>\! D^{*}_{\mathbf{y}} \!+ 3\delta, V_{\mathbf{{S}},\mathbf{{C}}} \in E_{L}(d)\right)  \leq \\[-6pt]
    & \ \ \ \ \ \ \ \ \ \ \ \ \ \ \ \ \ \ \ \ \ \ \ \ \ \ \ \ \ \ \ \ \ \ \ \ \ \ \ \ \ \ \prod\limits_{\mathbf{x}}(L_{\mathbf{x},\mathbf{y}}+1)^{|\mathcal{X}||\mathcal{Y}|} \exp\left(-L\left(\mathcal{D}^{*}_{\mathbf{y}}+ 3 \delta\right)\right),
    \end{aligned}
\end{equation}
since there are only a polynomial number of joint types. Then, again by \cite[Th. 11.1.4]{Thomas_Cover}, we observe that,
\begin{equation}
    \begin{aligned}
    &\mathbb{P}\left(\sum\limits_{\mathbf{x}}\mathbb{M}(\mathbf{x}|\mathbf{y})\mathcal{D}\left(V_{\mathbf{{S}},\mathbf{{C}}|\mathbf{x},\mathbf{y}}|| P(X|\mathbf{x}) \!\times Q(Y|\mathbf{y})\right) \!\leq\! D^{*}_{\mathbf{y}} \!+ 2\delta, V_{\mathbf{{S}},\mathbf{{C}}} \in E_{L}(d)\right) = \\[-5pt]
    & \ \ \sum\limits_{\substack{\\[-5pt]V^{'}_{\mathbf{x},\mathbf{y}} \in E_{{\mathbf{x},\mathbf{y}}}({d}_{\mathbf{x},\mathbf{y}}) \cap \mathcal{P}_{L_{\mathbf{x},\mathbf{y}}} \times \mathcal{Q}_{L_{\mathbf{x},\mathbf{y}}}: \\ \sum\limits_{\mathbf{x}}\mathbb{M}(\mathbf{x}|\mathbf{y})\mathcal{D}(V^{'}_{\mathbf{x},\mathbf{y}}||P(X|\mathbf{x})\times Q(Y|\mathbf{y}))\leq D^{*}_{\mathbf{y}} + 2\delta \\
    \sum\limits_{\mathbf{x},\mathbf{y}}\mathbb{M}(\mathbf{x},\mathbf{y}) \rho(x_{1},y_{1}) = \sum\limits_{\mathbf{x},\mathbf{y}}\mathbb{M}(\mathbf{x},\mathbf{y}) d_{\mathbf{x},\mathbf{y}}\leq d}} \prod\limits_{\mathbf{x}}\mathbb{P}(T_{\mathrm{y}}({V^{'}_{\mathbf{x},\mathbf{y}}})) \geq \frac{\exp(-L(\mathcal{D}^{*}_{\mathbf{y}}+2\delta))}{\prod\limits_{\mathbf{x}}(L_{\mathbf{x},\mathbf{y}}+1)^{|\mathcal{X}||\mathcal{Y}|}},
    \end{aligned}
\end{equation}
since for sufficiently large $L$, there exists at least one term in the summation, i.e., there exists a set of joint types $V^{'}_{\mathbf{x},\mathbf{y}}$ in $E_{L_{\mathbf{x},\mathbf{y}}}({d}_{\mathbf{x},\mathbf{y}}), \forall \mathbf{x} \in \mathcal{X}^{M}, \forall \mathbf{y} \in \mathcal{Y}^{M}$, such that,
\begin{equation}
    \sum\limits_{\mathbf{x}} \mathbb{M}(\mathbf{x}|\mathbf{y})\mathcal{D}\left(V^{'}_{\mathbf{x},\mathbf{y}}||P(X|\mathbf{x})\times Q(Y|\mathbf{y})\right)  \leq D^{*}_{\mathbf{y}} + 2\delta, \text{ and } \sum\limits_{\mathbf{x},\mathbf{y}}\mathbb{M}(\mathbf{x},\mathbf{y})  d_{\mathbf{x},\mathbf{y}}\leq d.
    \label{eq: asym_2_markov}
\end{equation}
Next, taking into account that the probability of one event is larger than or equal to the probability of the intersection, we have,
\begin{equation}
    \mathbb{P}\left(V_{\mathbf{{S}},\mathbf{{C}}} \in E_{L}({d})\right) \!\geq\!  \frac{\exp{\left(-L \left(D^{*}_{\mathbf{y}} \!+\! 2 \delta\right)\right)}}{\prod\limits_{\mathbf{x}}(L_{\mathbf{x},\mathbf{y}}+1)^{|\mathcal{X}||\mathcal{Y}|}} .\!\!
\end{equation}
By Bayes' law we get,
\begin{equation}
    \begin{aligned}
  \mathbb{P}\left(\sum\limits_{\mathbf{x}}\mathbb{M}(\mathbf{x}|\mathbf{y})\mathcal{D}\left(V_{\mathbf{{S}},\mathbf{{C}}|\mathbf{x},\mathbf{y}}|| P(X|\mathbf{x}) \!\times Q(Y|\mathbf{y})\right) \!>\! D^{*}_{\mathbf{y}} \!+ 3\delta \ \Big| \ V_{\mathbf{{S}},\mathbf{{C}}} \in E_{L}({d})\right) & \leq \\[-6pt]
  & \!\!\!\!\!\!\!\!\!\!\!\!\!\!\!\!\!\!\!\!\!\!\!\!\!\!\!\!\!\!\!\!\!\!\!\!\!\!\!\!\!\!\!\!\!\!\!\!\!\!\!\!\!\!\!\!\!\!\!\!\!\!\!\!\!\!\!\!\!\!\!\!\!\!\!\!\!\!\!\! \prod\limits_{\mathbf{x}}(L_{\mathbf{x},\mathbf{y}} + 1)^{2|\mathcal{X}||\mathcal{Y}|}\exp{\left(-L\delta\right)}.
    \end{aligned}
    \label{eq:Th_1}
\end{equation}
By the ``Pythagorean'' theorem \cite[Th. 11.6.1]{Thomas_Cover}, we have,
\begin{equation}
\begin{aligned}
    \!\!\mathcal{D}(V_{\mathbf{{S}},\mathbf{{C}}|\mathbf{x},\mathbf{y}}||V_{\mathbf{x},\mathbf{y}}^{*}) +& \mathcal{D}\!\left(V_{\mathbf{x},\mathbf{y}}^{*}||P(X|\mathbf{x})\!\times \!Q(Y|\mathbf{y})\right) \!\!\leq\! \mathcal{D}\!\left(V_{\mathbf{{S}},\mathbf{{C}}|\mathbf{x},\mathbf{y}}||P(X|\mathbf{x}) \! \times\! Q(Y|\mathbf{y})\right)\!.\!\!\!\!
    \end{aligned}
\end{equation}
 Hence, $\sum\limits_{\mathbf{x}}\mathbb{M}(\mathbf{x}|\mathbf{y})\mathcal{D}\left(V_{\mathbf{{S}},\mathbf{{C}}|\mathbf{x},\mathbf{y}}||P(X|\mathbf{x}) \times Q(Y|\mathbf{y})\right) \leq D^{*}_{\mathbf{y}} + 3\delta$ implies that,
\begin{equation}
    \sum\limits_{\mathbf{x}}\mathbb{M}(\mathbf{x}|\mathbf{y})\mathcal{D}(V_{\mathbf{{S}},\mathbf{{C}}|\mathbf{x},\mathbf{y}}||V_{\mathbf{x},\mathbf{y}}^{*}) \leq 3\delta.
\end{equation}
Finally, by the data processing inequality, we have,
\begin{equation}
    \!\!\sum\limits_{\mathbf{x}}\mathbb{M}(\mathbf{x}|\mathbf{y})\mathcal{D}(Q_{\mathbf{{C}}|\mathbf{x},\mathbf{y}}||Q^{*}({P}(X|\mathbf{x}),Q(Y|\mathbf{y}),{d}_{\mathbf{x},\mathbf{y}}^{*})) \leq\! \sum\limits_{\mathbf{x}}\mathbb{M}(\mathbf{x}|\mathbf{y})\mathcal{D}(V_{\mathbf{{S}},\mathbf{{C}}|\mathbf{x},\mathbf{y}}||V_{\mathbf{x},\mathbf{y}}^{*}),\!\!
\end{equation}
since, both are the respective $y$-marginals of the joint types. Hence, Theorem 1 follows from (\ref{eq:Th_1}) as desired. Furthermore, it is worth noting that,
\begin{equation}
    R(P(X|\mathbf{x}), Q(Y|\mathbf{y}), d_{\mathbf{x},\mathbf{y}}) = \min\limits_{V \in E_{{\mathbf{x},\mathbf{y}}}(d_{\mathbf{x},\mathbf{y}})} \mathcal{D}(V || P(X|\mathbf{x}) \times Q(Y|\mathbf{y})).
\end{equation}
Hence, by \cite{RD_Gray}, the minimum in (\ref{eq:min_average_div}) is achieved by adding the output-constrained rate-distortion functions at points of equal slopes in all co-ordinates, implying (\ref{eq:equal_slopes}).
\section*{Appendix C -
NTS Algorithm Alternating Minimization over Convex Sets \label{App_theorem_2}}
We can write the average rate-distortion function over all source-code cross sub-streams $\{\mathbf{s}_{\mathbf{x}},\mathbf{c}_{\mathbf{y}}\}$, that can be achieved by an output distribution with $M$-th order Markov property, as an average of double minimization over \textit{convex} sets, i.e.,
\begin{equation}
    \overline{R}({d}) = \sum\limits_{\mathbf{x},\mathbf{y}} \mathbb{M}(\mathbf{x})\mathbb{M}^{*}(\mathbf{y}|\mathbf{x}) R_{\mathbf{x},\mathbf{y}}(P(X|\mathbf{x}),d^{*}_{\mathbf{x},\mathbf{y}}),
\end{equation}
\begin{equation}
    \overline{R}(d)= \min\limits_{Q(Y|\mathbf{y})} \min\limits_{\substack{\mathbb{{M}}(\mathbf{y}|\mathbf{x})\\
    {d}_{\mathbf{x},\mathbf{y}}, V_{\mathbf{x},\mathbf{y}}}} \sum\limits_{\mathbf{x},\mathbf{y}} \mathbb{M}(\mathbf{x}) \mathbb{M}(\mathbf{y}|\mathbf{x}) \ \mathcal{D}\left(V_{\mathbf{x},\mathbf{y}} \ \big|\big| \  P(X|\mathbf{x}) \times Q(Y|\mathbf{y})\right).
    \label{eq:R_markov_pd}
\end{equation}
such that $\mathbb{M}(\mathbf{y}|\mathbf{x})$, $d_{\mathbf{x},\mathbf{y}}$ and $V_{\mathbf{x},\mathbf{y}}$ satisfy conditions in (\ref{eq:conditions_1}). For ease of notation, let,
\begin{equation}
    \overline{D} = \sum\limits_{\mathbf{x},\mathbf{y}} \mathbb{M}(\mathbf{x}) \mathbb{M}(\mathbf{y}|\mathbf{x}) \ \mathcal{D}\left(V_{\mathbf{x},\mathbf{y}} \ \big|\big| \  P(X|\mathbf{x}) \times Q(Y|\mathbf{y})\right).
\end{equation}
It should be noted that all constraints on optimization variables in (\ref{eq:R_markov_pd}) are convex, and it is easy to verify that this yields convex sets. Hence, for a fixed set of joint distributions $\{V_{\mathbf{x},\mathbf{y}}\}$, the reproduction conditional distribution $Q(Y|\mathbf{y})$ which minimizes the average divergence $\overline{D}$ is the $y$-marginal of $\sum_{\mathbf{x}} \mathbb{M}(\mathbf{x}|\mathbf{y}) V_{\mathbf{x},\mathbf{y}}$ on $\mathcal{Y}$. On the other hand, for a fixed set of conditional distributions $\{Q(Y|\mathbf{y})\}$ and distortion constraint ${d}$, the joint distribution which minimizes $\overline{D}$ under constraints in (\ref{eq:conditions_1}) will induce $Q(Y|\mathbf{y})=\sum_{\mathbf{x}} \mathbb{M}^{*}(\mathbf{x}|\mathbf{y}) Q^{*}(P(X|\mathbf{x}), Q(Y|\mathbf{y}), d^{*}_{\mathbf{x},\mathbf{y}})$. By the results of Theorem 1, the recursion in (\ref{eq:NTS_markov_recursion}) performs exactly this minimization over the convex sets. It should be noted that the distance in the alternating minimization is measured by divergence. Hence, by \cite[Th. 3]{Alternating_Min}, the sequences of divergences and distributions will converge to the minimum divergence, i.e., $\overline{R}({d})$, and the corresponding reproduction distributions.

\bibliographystyle{unsrt}  
\bibliography{templateArxiv}

\end{document}